\documentclass[twocolumn,aps,reprint,pra]{revtex4-2}
\usepackage[english]{babel}
\usepackage{amsmath,amsfonts,amsthm,bm,braket}

\usepackage{graphicx}

\usepackage{float}
\usepackage{physics}
\usepackage{tikz}
\usepackage[normalem]{ulem}
\usetikzlibrary{trees}
\DeclareUnicodeCharacter{2212}{-}

\usepackage{bbold}
\usepackage{comment}
\usepackage{adjustbox}
\usepackage{mathtools}
\usepackage{enumerate}
\usepackage{svg}
\usepackage{relsize}
\usepackage{subcaption}

\usepackage{natbib}

\begin{document}

\title{Classification using quantum kernels in a radial basis function network}%

\author{E. Micklethwaite}
\affiliation{QinetiQ, Malvern Technology Centre, St Andrew’s Road, Malvern, WR14 3PS, UK }%

\author{A. Lowe}%
 \email{adam.lowe@qinetiq.com}
\affiliation{QinetiQ, Malvern Technology Centre, St Andrew’s Road, Malvern, WR14 3PS, UK }%

\date{\today}

\begin{abstract}Radial basis function (RBF) networks are expanded to incorporate quantum kernel functions enabling a new type of hybrid quantum-classical machine learning algorithm. Using this approach, synthetic examples are introduced which allow for proof of concept on interpolation and classification applications. Quantum kernels have primarily been applied to support vector machines (SVMs), however the quantum kernel RBF network offers potential benefit over quantum kernel based SVMs due to the RBF networks ability to perform multi-class classification natively compared to the standard implementation of the SVM.
\end{abstract}

\maketitle

\section{Introduction}
The development of quantum computing in the past decade has led to the suggestion of several potential applications which may provide benefit over classical computing \citep{doi:10.1126/science.abn7293,google1,google2,Huggins_2019}. One such avenue is the advent of quantum machine learning, which traditionally combines state-of-the-art classical machine learning algorithms with quantum mechanical approaches \citep{QML_rev,VQA,Q_Reinf_learn, Q_CNN}. These types of algorithms are typically split into two branches. One approach is to develop an entirely quantum algorithm to undertake common machine learning algorithms \citep{Q_PCA}. A separate approach is to use quantum computers for a given step within a classical machine learning algorithm \citep{PARK2020126422}. This second approach is commonly referred to as a hybrid quantum-classical algorithm, due to the requirement of both quantum and classical hardware to perform the algorithm. 

One of the most developed hybrid approaches is the quantum support vector machine (QSVM) \citep{PhysRevLett.113.130503,natQSVM,qsvm_complexity}. The SVM is a binary classifier which makes use of the 'kernel trick', whereby a non-linear system in a given dimensional space can be mapped to a higher dimensional space where the system can be solved linearly \citep{svm_ref}. One of the key benefits of this trick is that the explicit evaluation of the projection of the datapoints in the higher dimensional space is not required, only the result of the inner product of a chosen feature map. Given the inherent linearity of quantum mechanics, the computation of an inner product in a high dimensional space offers the possibility for entanglement to provide possible benefit. QSVMs have been utilised for a variety of different applications, including regression, classification and solving differential equations \citep{PhysRevA.107.032428, kernel_class}. There have also been several experimental demonstrations of quantum classifiers being experimentally realised on real quantum hardware \citep{PhysRevLett.114.140504,qsvm_class_photonic,qsvm_class_trapped_ion}.

Whilst SVMs offer a practical, mathematically tractable algorithm for undertaking binary classification, there are other approaches which have not been investigated from a quantum perspective. One of the alternate approaches is the radial basis function (RBF) network \citep{Broomhead1988MultivariableFI}. This is an analytically tractable artificial neural network which uses one hidden layer for the computation of the basis functions. It should be highlighted that the RBF network is distinct to what is commonly referred to in the literature as the RBF kernel, as the RBF kernel is a specific choice of basis function, which is commonly used for SVMs. Given the RBF kernel is a Gaussian function, it shall henceforth be referenced as a Gaussian kernel. 

RBFs were originally introduced as a function interpolator, where parameters are fitted according to the input and output data. However, they have also been used extensively for prediction, regression, visualisation and classification purposes \citep{shakespeare,PhysRevD.104.076011,6795628,TIPPING1998211}. A key benefit of using RBFs is that they can natively perform multi-class classification without the requirement of further developing the algorithm to do so, unlike SVMs \citep{991427,svm_ref1}. By using the kernel trick, RBFs are also well suited to solving non-linear problems as it is only the outcome of the kernel functions that are evaluated, rather than the individual functions required for the inner product. In order to fit the parameters, a distance metric in the high dimensional space is computed, between the input datapoints and chosen `centre' datapoints. These chosen datapoints, or centres, are introduced allowing for the calculation of the dissimilarities between each input datapoint and the corresponding centre. Ensuring that the number of centres is less than the number of input datapoints avoids overfitting the RBF model to the data. Further mathematical details are provided in Sec. \ref{sec2}.

Given the potential benefit of using quantum kernels in machine learning, this paper will look to develop a radial basis function network which incorporates quantum kernels, henceforth referred to as a quantum radial basis function (Q-RBF). It should be highlighted that a quantum radial basis function network has previously been developed through a separate approach \citep{PhysRevA.102.042418}. However, it attempts to use quantum mechanics to describe the whole algorithm, rather than a hybrid quantum-classical approach. In the approach presented here, the quantum computation is only relevant for computing the kernel metric, in a similar fashion to the QSVM. A hybrid quantum-classical approach has been developed previously \citep{ZHOU2025129254}, however this paper extends upon the mathematical approach, the data encoding, and considers both interpolation and classification.

The structure of this paper is as follows: In Sec. \ref{sec2}, the mathematical formulation of the RBF is detailed, in addition to highlighting the construction of the quantum kernel and how it is embedded into the existing classical algorithm; Sec. \ref{interpolation} provides proof of concept of this approach for interpolating several well-known functions with comparisons against existing approaches; Sec. \ref{classification} demonstrates the capability of the Q-RBF network at classifying both images and generated datasets; Finally Sec. \ref{discussion} discusses future applications and implementations of this approach.

\section{Q-RBF Network Derivation} \label{sec2}

The classical radial basis function (C-RBF) is defined as
\begin{equation}
\label{class_rbf_def}
    {\bf{f}}({\bf{x}}_i) = \sum_{j=1}^{n} {\boldsymbol{\beta}}_j \phi_j ( \norm{{\bf{x}}_i - {\bf{y}}_j}),
\end{equation}
where ${\bf{x}}_i \in \{{\bf{x}}_1, {\bf{x}}_2, \dots , {\bf{x}}_m \}$ represents the input data, and ${\bf{y}}_j \in \{{\bf{y}}_1, {\bf{y}}_2, \dots , {\bf{y}}_n \}$ represents the chosen centres in the high dimensional space, where $m,n$ represent the number of datapoints and centres respectively. ${\boldsymbol{\beta}}_j$ are coefficients to be determined, and ${\bf{f}}({\bf{x}}_i)$ is the function output. Formally, this provides a mapping where ${\bf{f}}({\bf{x}}_i):\mathbb{R}^{P} \rightarrow \mathbb{R}^{q}$, where $P$ represents the dimensionality of the input data, and $q$ is the dimensionality of the output data. Additionally, $\phi ( \norm{{\bf{x}}_i - {\bf{y}}_j})$ is the kernel function, which is a distortion of the dissimilarity measure between input data points and chosen basis function centres. For the classical RBF, the kernel metric measures the Euclidean distance between these two points in this data space, since $\norm{\dots}$ denotes the Euclidean norm. Therefore, Eq. (\ref{class_rbf_def}) can be written in matrix form as
\begin{equation}
\label{rbf_matrix}
    \hat{\boldsymbol{\phi}} \hat{\boldsymbol{\beta}} = \hat{\bf{f}},
\end{equation}
where the hat denotes matrix representation, and each element of the matrix $\hat{\boldsymbol{\phi}}$ is a kernel function represented by $\phi ({\bf{x}}_i,{\bf{y}}_j)$. Dimensionally, $\hat{\boldsymbol{\phi}} \rightarrow \mathbb{R}^{m \times n}$ and $\hat{\boldsymbol{\beta}} \rightarrow \mathbb{R}^{n \times q}$. Explicitly, $\hat{\boldsymbol{\phi}}$ is given by
\begin{equation}
\label{phimat}
    \hat{\boldsymbol{\phi}} =  \begin{pmatrix} \phi_1 ({\bf{x}}_1,{\bf{y}}_1) & \dots & \phi_n ({\bf{x}}_1,{\bf{y}}_n) \\ \vdots & \ddots & \vdots \\ \phi_1 ({\bf{x}}_m,{\bf{y}}_1) & \dots & \phi_n ({\bf{x}}_m,{\bf{y}}_n)  \end{pmatrix}.
\end{equation}
In the classical RBF network, different kernel functions can be used including linear kernels ($z$), Gaussian kernels ($\exp (-\gamma z^2)$), and splines ($z\log z$), where $z=\norm{{\bf{x}}_i - {\bf{y}}_j}$ \citep{374353}. An additional subtlety occurs due to the ability to choose different kernel functions for each $j$. However, in most situations the kernel function is chosen to be consistent for all $j$, e.g. a linear kernel for all $j$. 

Eq. (\ref{rbf_matrix}) can be solved to find values for $\hat{\boldsymbol{\beta}}$ which are fitted parameters which relate the input data to the output function. At this stage, it is important to clarify some key aspects regarding the classical RBF network. In order to find $\hat{\boldsymbol{\beta}}$, the Moore-Penrose pseudoinverse is often used such that $\hat{\boldsymbol{\beta}} = \hat{\boldsymbol{\phi}}^{+}\hat{\bf{f}} \equiv \hat{\boldsymbol{\beta}}^{*} $ as generally $\hat{\boldsymbol{\phi}}$ is a non-square matrix. This is because if the number of centres are chosen to match the number of datapoints, then this leads to overfitting the $\hat{\boldsymbol{\beta}}$ parameters to the input data as this results in strict interpolation. This leads to issues when trying to model unseen data, due to overfitting the model to the training data, as is discussed later in this manuscript. Also, whilst ${\bf{x}}_i$ and ${\bf{y}}_j$ are generally not scalars, examples including the interpolation of 1D time series can yield a scalar $x_i$ and $y_j$. However, for classification problems (especially multi-class classification), it is usually necessary for the elements of ${\bf{x}}_i$ and ${\bf{y}}_j$ to be non-scalar. Encoding extra features of datasets into ${\bf{x}}_i$ enables multi-class classification to be performed natively using a RBF network, in contrast to the commonly used SVM approach.

Once $\hat{\boldsymbol{\beta}}^{*}$ is found, this can be used with unseen test data for inference purposes for both interpolation, prediction, and classification. This manuscript will focus on understanding how effective the found set of $\hat{\boldsymbol{\beta}}^{*}$ is for inference using both classical and quantum kernels.
 
Given the RBF network is a kernel-based approach for interpolation and classification problems, it provides an ideal opportunity to understand how quantum kernels could be incorporated into the theory, and whether considering overlap of wavefunctions may be a preferential approach compared to Euclidean distances. Subsequently, the use of quantum kernels is proposed within a RBF network. The significant difference between the two approaches is how the kernel metric is constructed, as now the quantum kernel denoted by $\kappa ({\bf{x}}_i,{\bf{y}}_j)$ is given by $\kappa ({\bf{x}}_i,{\bf{y}}_j) = \abs{\braket{\psi({\bf{y}}_j)}{\psi({\bf{x}}_i)}}^2$. It should be highlighted at this stage that this quantum kernel metric could be incorporated into a classical kernel function, as in the classical scenario.

The main question related to quantum kernels is how to encode the classical data into the quantum wavefunction. Below, one approach is outlined which details how the input data is encoded onto the Bloch sphere through rotation matrices, and then entangled through either a CNOT gate, or a random Haar unitary matrix \citep{Mele2024introductiontohaar,PRXQuantum.5.010309}. 

Initially, the input data needs to be encoded onto the Bloch sphere through the $RX$ gate (where the standard factor of 1/2 inside the trigonometric arguments is neglected) such that,
\begin{equation}
\label{rxgate}
\begin{split}
    \big(\tilde{R}_x ({\bf{x}}_k) \big)_p &= e^{-i {\boldsymbol{\sigma}_x} {{\bf{x}}_k} (p)} \\&= {\boldsymbol{\sigma}_0} \cos {{\bf{x}}_k} (p) 
    - i {\boldsymbol{\sigma}_x} \sin {{\bf{x}}_k} (p),
\end{split}
\end{equation}
where $\boldsymbol{\sigma}_x$ is the $X$-Pauli matrix, $\boldsymbol{\sigma}_0$ is the identity matrix, and ${{\bf{x}}_k}(p)$ represents an element from the input dataset, where $p$ denotes the index for the element from the input dataset vector. It should also be noted that ${{\bf{x}}_k}(p)$ is rescaled such that ${{\bf{x}}_k}(p)\rightarrow \alpha {{\bf{x}}_k}(p)$ to ensure that each datapoint maps to a different part of the Bloch sphere, where $\alpha$ is typically chosen to be approximately the inverse of the maximum value of the dataset such that ${{\bf{x}}_k}(p) \in [0,1]$, when all input data points are postive semi-definite. However, subtleties can be introduced which take advantage of the periodicity of the trigonometric functions, in addition to adequately taking into account negative data inputs. One potential approach is to normalise the data to account for this. Whilst this is not undertaken in this paper, other approaches are discussed in Secs. \ref{interpolation} and \ref{classification}. Using Eq. (\ref{rxgate}) each datapoint is encoded onto one qubit, where the point on the surface of the Bloch sphere represents the encoded datapoint. However, in order to allow quantum effects within a single datapoint, each individual feature is encoded onto two qubits such that
\begin{equation}
\label{R_encoding}
    \big(R_x ({\bf{x}}_k)\big)_p = \big(\tilde{R}_x ({\bf{x}}_k) \big)_p \otimes \big(\tilde{R}_x ({\bf{x}}_k) \big)_p.
\end{equation}
This approach can also be used for encoding the basis function centres onto the qubits as well. Generally ${\bf{x}}_k$ is a vector, so each scalar element within the datapoint ${\bf{x}}_k$ must be encoded through the $p$-index. To do this generically, the input datapoint dimension $P$ is used. For example, when $P=3$ and $k=1$, ${\bf{x}}_1 \in \{x_1, x_2, x_3\}$ and ${\bf{x}}_k (p=1) = x_1$. When $P=1$, the input datapoint has a length of 1 and yields a scalar. Subsequently, the general form of Eq. (\ref{R_encoding}) to account for each element within the datapoint is
\begin{equation}
     R_{x} ({\bf{x}}_k) = \bigotimes_{p = 1}^{P} \big(R_x ({\bf{x}}_k)\big)_p,
\end{equation}
where $p$ spans all elements within given input datapoints. 
\begin{figure*}[ht]
\begin{subfigure}{.5\linewidth}
  \includegraphics[width=\linewidth]{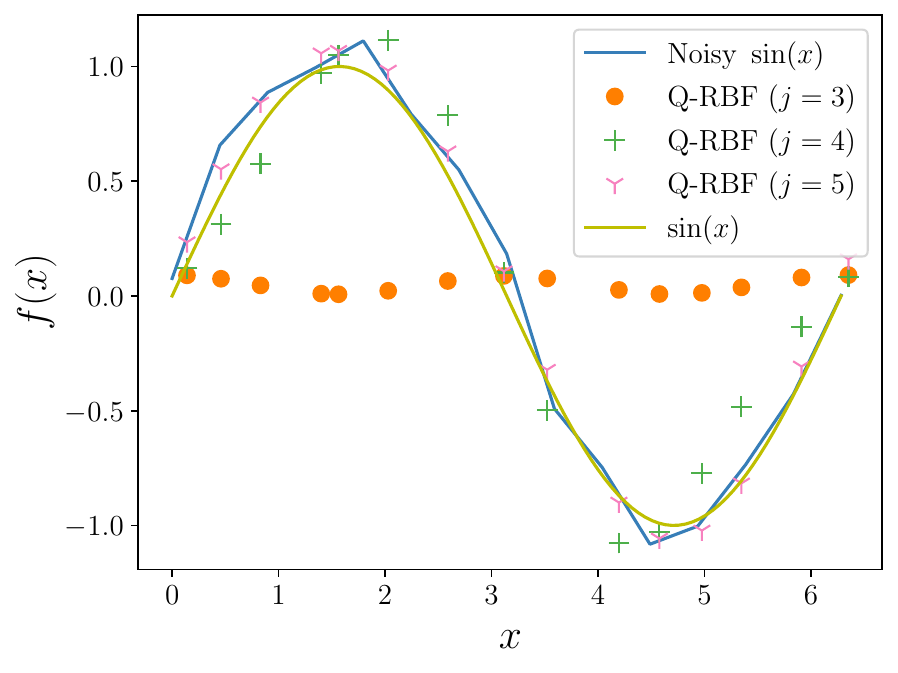}
  \caption{}
\end{subfigure}\hfill
\begin{subfigure}{.5\linewidth}
  \includegraphics[width=\linewidth]{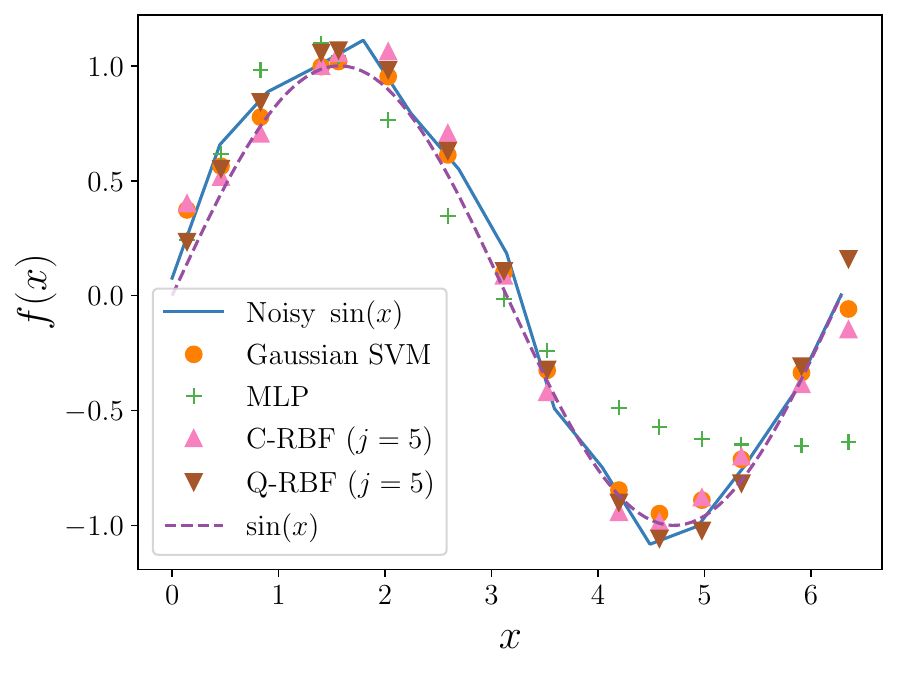}
  \caption{}
\end{subfigure}
    \caption{(a) Interpolation of a sine wave with an increasing number of centres. It is observed that increasing the number of data centres results in an improved fit. This is consistently seen throughout this paper. (b) Interpolation of a sine wave using Gaussian SVM, MLP, C-RBF and Q-RBF models. This subfigure demonstrates the RBF's ability to successfully interpolate data points, along with the Gaussian SVM. The MSE for this plot is shown in Table 1.}
    \label{fig:Interpolator_Sin}
\end{figure*}
Using this resulting rotation matrix, an entangling unitary gate can be used which entangles the qubits (and thus the features within the datapoints) with themselves. When $x_k$ is a scalar, a CNOT gate can be used as this entangles two qubits. However, when ${{\bf{x}}_k}$ is non-scalar and contains multiple scalar features (e.g. $P\geq 2$), this requires more than two qubits to be entangled. Therefore, a random Haar unitary is chosen which generates some entanglement between a larger number of qubits (e.g. for three qubits, an $8\times 8$ Haar unitary matrix is required).
Therefore, the encoded wavefunction is defined by
\begin{equation}
    \ket{\psi({\bf{x}}_i)} = \hat{U} R_{x} ({{\bf{x}}_i}) \ket{0}^{\otimes 2P},
\end{equation}
where $\hat{U}$ denotes the entangling matrix (CNOT in the two qubit case, a Haar unitary for greater than two qubits), $i$ is an integer which runs from $1, 2, \dots, m-1, m$, where $m$ is the length of the dataset, and all other terms are as defined previously. The factor of two in the tensor product of the $\ket{0}$ state is due to each element within the datapoint being encoded onto two qubits. Subsequently, this resultant wavefunction allows $\hat{\boldsymbol{\phi}}$ to be computed using kernel metrics which are constructed from the overlap of training data points and the chosen centres through quantum wavefunctions, such that
\begin{equation}
    \hat{\boldsymbol{\phi}} =  \begin{pmatrix} \kappa_Q ({\bf{x}_1},{\bf{y}_1}) & \dots & \kappa_Q ({\bf{x}_1},{\bf{y}_n}) \\ \vdots & \ddots & \vdots \\ \kappa_Q({\bf{x}_m},{\bf{y}_1}) & \dots & \kappa_Q({\bf{x}_m},{\bf{y}_n})  \end{pmatrix}.
\end{equation}
At this stage, this is assuming that each element of $\hat{\boldsymbol{\phi}}$ is a linear kernel function, such that $\kappa ({\bf{x}}_i,{\bf{y}}_j) = \phi(\abs{\braket{\psi({\bf{y}}_j)}{\psi({\bf{x}}_i)}}^2) = \abs{\braket{\psi({\bf{y}}_j)}{\psi({\bf{x}}_i)}}^2$. However, whilst the wavefunctions and their inner products would be computed on quantum hardware, $\hat{\boldsymbol{\phi}}$ is a matrix of classical values, and the RBF network can be solved in the same way as when using classical kernels. It would be interesting to consider how different choices of kernel functions may lead to improved results, e.g. understanding the effect of a Gaussian kernel function compared to a linear kernel function. However, this is out of scope for this paper, and is left for future work. 
The following section details how this approach performs on synthetic examples in simulation. Future work will involve implementing this approach using quantum simulators and on quantum hardware.

The testing undertaken throughout this paper involved using publicly available Python packages for comparative purposes. The \textbf{svm.SVR} and \textbf{svm.SVC} functions from the \textbf{sklearn} package were used for SVM testing, for regression and for classification with the kernel being selected as either Linear or Gaussian to understand the effect of the kernel. For the multilayer neural network, the \textbf{MLPRegressor} and \textbf{MLPClassifier} functions from the \textbf{sklearn} package were used for regression and classification testing, with two hidden layers and 50 neurons in each layer. The activation function chosen was $\tanh$, and the optimiser was \textbf{Adam}. The implementation of the classical RBF was based on publicly available code \cite{classicalRBF}. For the classical RBF, centres are chosen using \textbf{K-Means} clustering, as in \cite{classicalRBF}. For the quantum RBF, the centres are chosen via uniform sampling (\textbf{linspace}) for the interpolation testing and via random sampling from a Gaussian function (\textbf{random.gauss}) for the classification testing.

\section{Interpolation} \label{interpolation}

Given RBF networks were originally introduced in the context of interpolation, determining whether the quantum RBF network can interpolate a range of different functions is necessary to understand the efficacy of the algorithm. Mathematically, this problem can be formulated as, given a set of $m$ input datapoints (vectors), such that ${\bf{x}}_i$, where $i=1,2,\dots,m$ and $m$ output datapoints, ${\bf{f}}_i$, where $i=1,2,\dots,m$, find a function which maps the interpolation conditions such that $s({\bf{x}}_i) = {\bf{f}}_i$, where $s$ is the interpolating function.

Therefore, attempting to solve this interpolation problem on well-understood examples should provide an ideal framework for testing and determining whether a quantum RBF network performs comparably to its classical counterparts. All the interpolated tested models are fitted to noisy data to better represent real data modelling, given all realistic datasets have underlying noise associated with them. However, for each of the examples considered, the generating function without noise is also plotted for comparison to determine the model fitting on the unperturbed data.

\subsection{Sinusoidal Function} 
An initial test example to consider is a sine wave, as it is a non-linear but periodic function. Formally, we have the function $f(x) =\sin x$ 
where the interpolation is over the region $x \in [0,2\pi]$. For this problem, $P=q=1$ given it is a univariate time series, with a single class. Therefore the input and output datapoints are scalar. However, given the quantum RBF network needs to be trained on samples from the distribution space of input datapoints $x_i$ which map to the distribution space of $f(x_i)$, a finite number of samples needs to be chosen. For this example, 15 samples are chosen in order to train the coefficients in $\hat{\boldsymbol{\beta}}$. Alongside this, a variety of centres are chosen to estimate the overlap between the centres and the data points in the high dimensional Hilbert space. These centres are chosen from the data space which generated the training data, equally spaced over the interval range. Based on the found coefficients, the resulting function ($\hat{\bf{f}}^{*} = \hat{\boldsymbol{\phi}}\hat{\boldsymbol{\beta}}^{*}$) enables a function which interpolates the data generator over the same interval, but with different testing datapoints. The test datapoints have been randomly sampled over the interval for the training data. The results detailing the success of the interpolation are found in Table \ref{MSE} and Figure \ref{fig:Interpolator_Sin}. Table \ref{MSE} shows a comparison of the mean squared error for the interpolation of numerous machine learning models, including both quantum and classical RBF networks, linear and Gaussian SVMs and a multilayer neural network (MLP).
\begin{figure*}[ht]
\begin{subfigure}{.5\linewidth}
  \includegraphics[width=\linewidth]{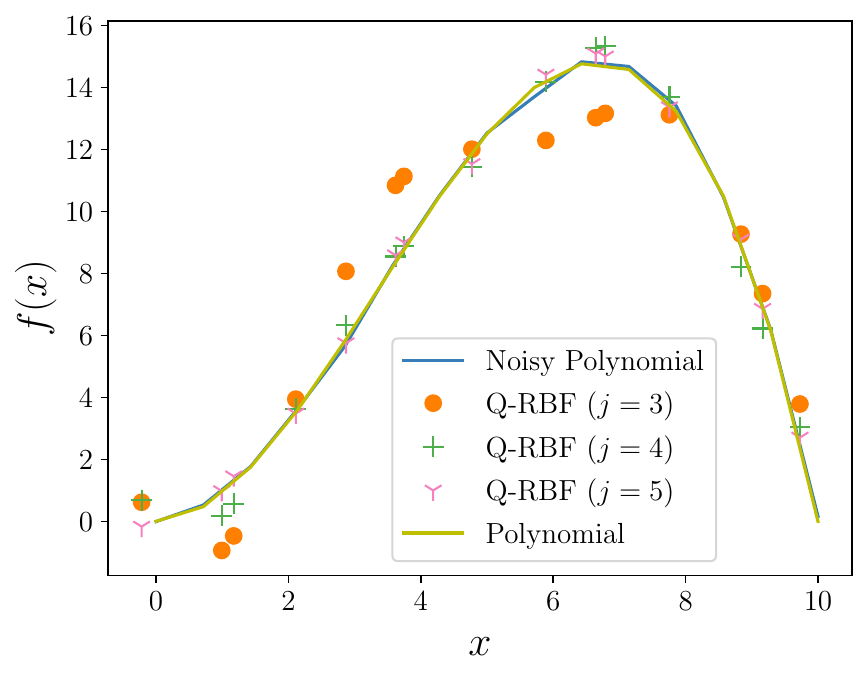}
  \caption{}
\end{subfigure}\hfill
\begin{subfigure}{.5\linewidth}
  \includegraphics[width=\linewidth]{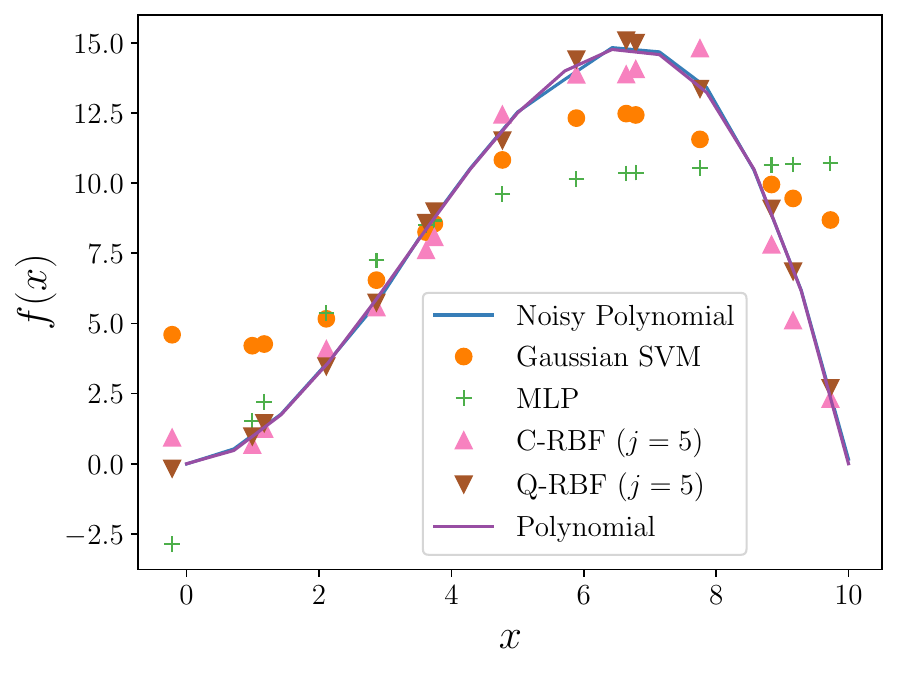}
  \caption{}
\end{subfigure}
    \caption{(a) Interpolation of the polynomial function in Eq. (\ref{polyeq}) with an increasing number of centres. Similar trends are observed to Fig. 1 highlighting this model can be generalised. (b) Interpolation of the polynomial function in Eq. (\ref{polyeq}) using Gaussian SVM, MLP, C-RBF and Q-RBF models. As before, the RBF's and Gaussian SVM are interpolating the data well. The MSE for this plot is shown in Table 1.}
        \label{fig2}
\end{figure*}

\begin{figure*}[ht]
\begin{subfigure}{.5\linewidth}
  \includegraphics[width=\linewidth]{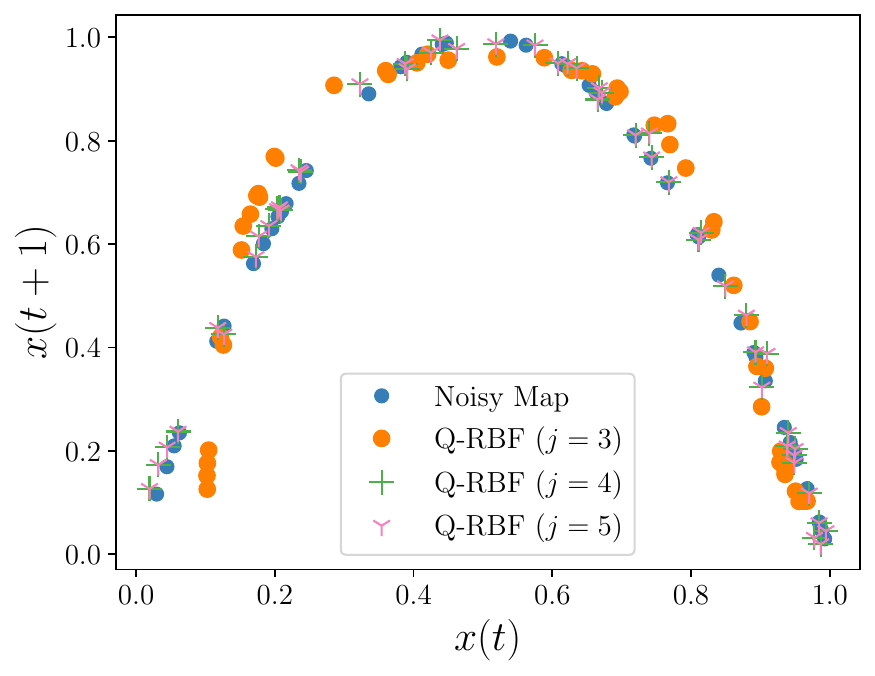}
  \caption{}
\end{subfigure}\hfill
\begin{subfigure}{.5\linewidth}
  \includegraphics[width=\linewidth]{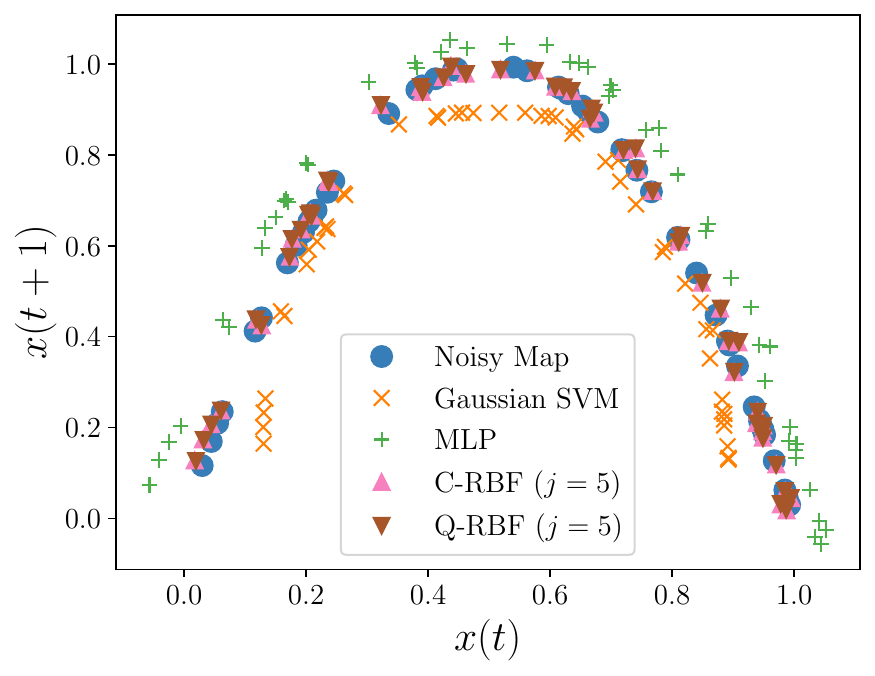}
  \caption{}
\end{subfigure}
    \caption{(a) Interpolation of the Logistic map in Eq. (\ref{logisticeq}) with an increasing number of centres. In this subfigure, the axes are $x(t)$ against $x(t+1)$ to better demonstrate the fit, given the time series of the function appears as $\delta$-correlated random noise. (b) Interpolation of the Logistic map in Eq. (\ref{logisticeq}) using Gaussian SVM, MLP, C-RBF and Q-RBF models. Similar trends to before are observed, highlighting the consistency with which the RBF's and SVM's interpolate successfully. The MSE for this plot is shown in Table 1.}
\end{figure*}

A key consideration with the interpolation of the sine function was the need to encode periodicity into the rotation matrix. This is due to the fact that several input values map to the same point on the Bloch sphere. Subsequently, the mapping $x_i \rightarrow \alpha x_i$ was performed where $\alpha = \pi/x_m$. This is an interesting feature of this data encoding as the periodicity of the rotation matrix can be used to enhance the quality of the fit. Understanding the optimal choice of this hyperparameter, and the resultant effect on accuracy for each given dataset is an interesting question which should be addressed in future work. \newline

\subsection{Polynomial Function}
To consider different types of functions other than trigonometric, it is useful to consider polynomial functions as well. The polynomial considered here is
\begin{equation}
\label{polyeq}
    f(x) = x^2 -0.1x^3,
\end{equation}
where $x \in [0,10]$, which captures both quadratic and cubic behaviour. Whilst again this is a trivial function to interpolate, it demonstrates that the quantum RBF also has the capability to fit to non-periodic functions. This problem also has parameter values $P=q=1$ for the same reasons as the sine wave example, and 15 datapoints were chosen for the interpolation, and the centres were chosen equally spaced over the input data range. The implementation of the training and testing was conducted as before with the fitting being performed on the noisy polynomial data, and the input testing data being perturbed to determine the quality of fit. The quantum RBF performs comparatively well based off the naive implementations of other commonly used techniques for performing interpolation. The results for this are visualised in Fig. \ref{fig2}, demonstrating that the RBF models fit the underlying generating model accurately.

\subsection{Logistic Map}
The third example is the logistic map, as it can generate chaotic behaviour for certain choices of coefficients. The output time series has the correlation structure of a random time series, without prior knowledge of the generating function. The logistic map is defined by
\begin{equation}
\label{logisticeq}
    x(t+1) = r x(t)(1-x(t)),
\end{equation}
where $t$ is an integer dummy variable, $r=4$, $x(0)=0.3$, and $x_0 \in [0,1]$. In order to interpolate for chaotic behaviour $r$ is chosen to be 4, as this is above a threshold where chaotic behaviour emerges \citep{log_map}. The function is bounded between $0$ and $1$ for all values of $t$. The key result here is that it appears that the quantum RBF can {\it{predict}} random time series data. This is of course not the case, but it highlights the quantum RBF's ability to interpolate the generating model of the data, rather than trying to fit exactly to the data points themselves. Here, $P=q=1$ as before.

\begin{table}
    \centering \caption{Out-of-sample Mean Square Error for Noisy Interpolation Models} \label{MSE}
\begin{tabular}{ c | c | c | c }
\hline
 & $\sin x$ & Polynomial & Logistic Map\\
\hline \hline
 Q-RBF $j=3$ & 0.539 & 2.95 & 0.000869 \\ 
 \hline
 Q-RBF $j=4$ & 0.0473 & 1.38 & 0.0000861 \\  
 \hline
 Q-RBF $j=5$ & 0.0128 & 0.887 & 0.0000861 \\
 \hline
 Classical RBF $j=5$ & 0.0241 & 1.59 & 0.0000877\\
 \hline 
 Linear SVM & 0.257 & 30.7 & 0.00426\\
 \hline
 Gaussian SVM & 0.0167 & 10.0 & 0.00228\\
 \hline
 MLP & 0.076 & 14.5 & 0.00233
\end{tabular} 
\end{table}

\begin{figure}
    \centering
    \includegraphics[width=\linewidth]{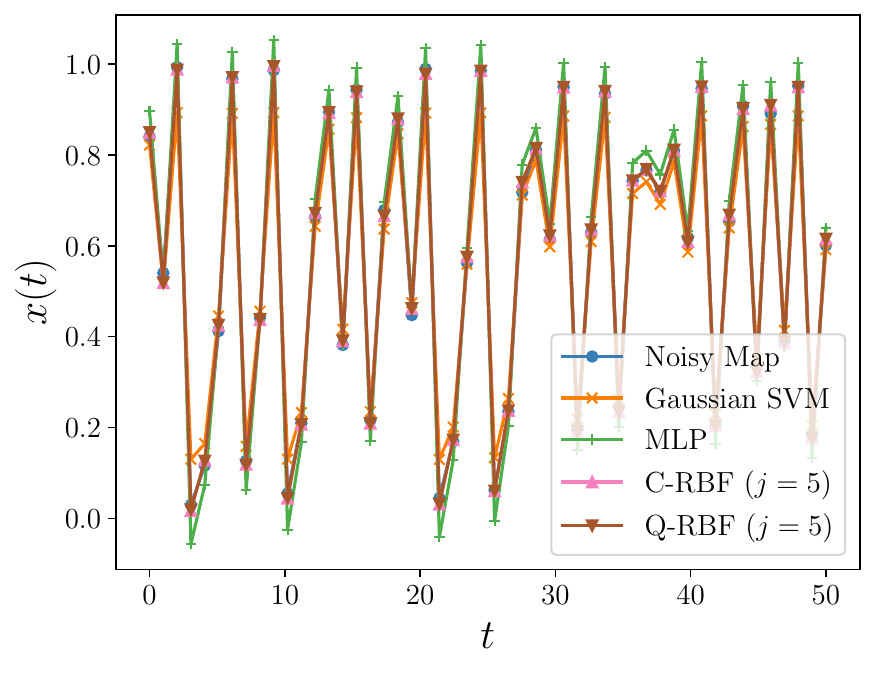}
    \caption{Interpolation of the Logistic map in Eq. (\ref{logisticeq}), viewed as a timeseries. Despite appearing as random data, the RBF models are able to infer the underlying generating model.}
    \label{logistic map timeseries}
\end{figure}

\begin{figure}
    \centering
    \includegraphics[width=\linewidth]{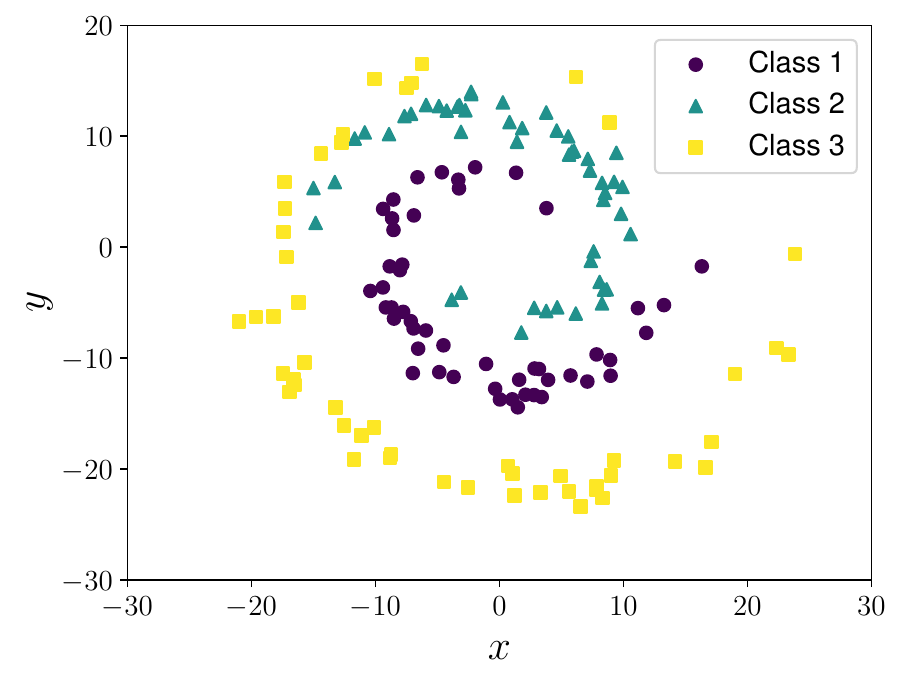}
    \caption{This highlights the 3 classes of spirals which are generated using Eq. (\ref{spiral_eq}). Determining which class each datapoint is generated from is the aim of the classification task.}
    \label{spiral dataset}
\end{figure}

\begin{figure*}[ht]

\begin{subfigure}{.5\linewidth}
  \includegraphics[width=\linewidth]{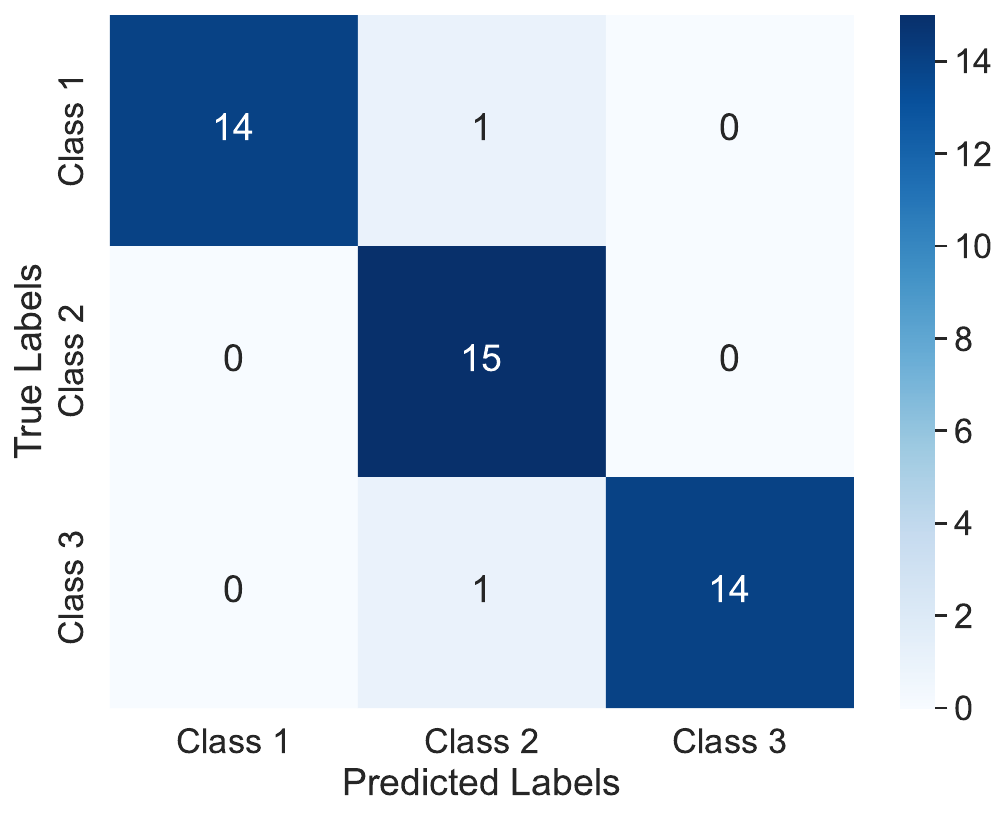}
  \caption{}
\end{subfigure}\hfill
\begin{subfigure}{.5\linewidth}
  \includegraphics[width=\linewidth]{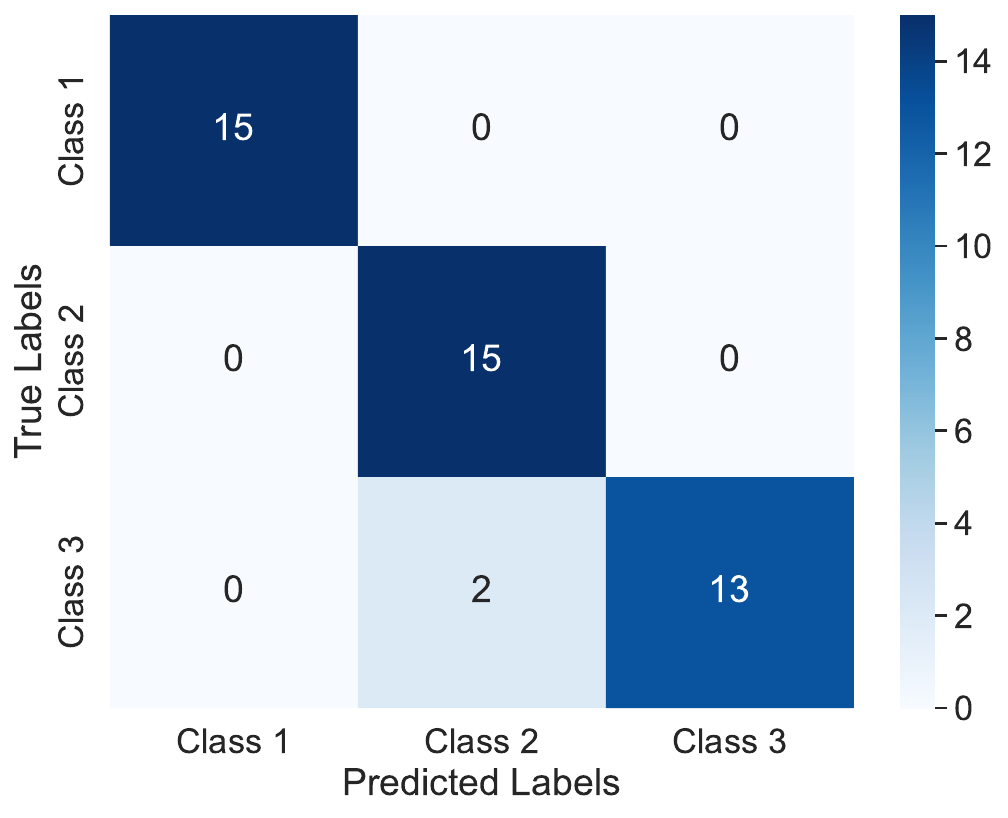}
  \caption{}
\end{subfigure}

\medskip 
\begin{subfigure}{.5\linewidth}
  \includegraphics[width=\linewidth]{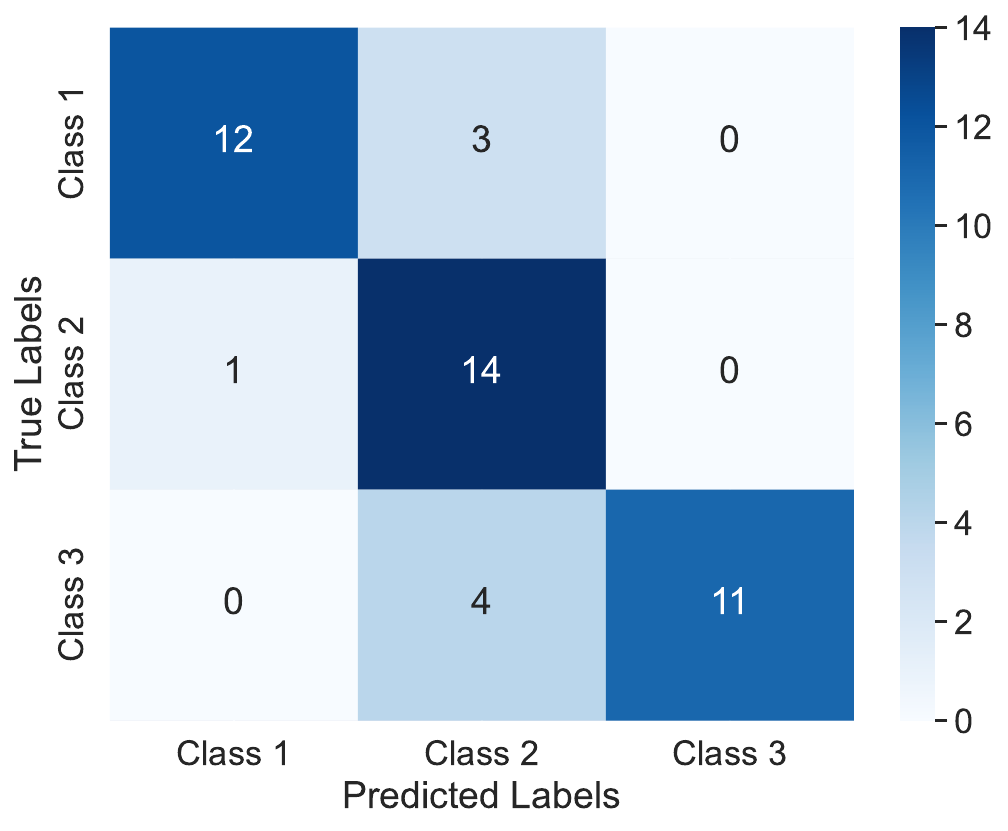}
  \caption{}
\end{subfigure}\hfill 
\begin{subfigure}{.5\linewidth}
  \includegraphics[width=\linewidth]{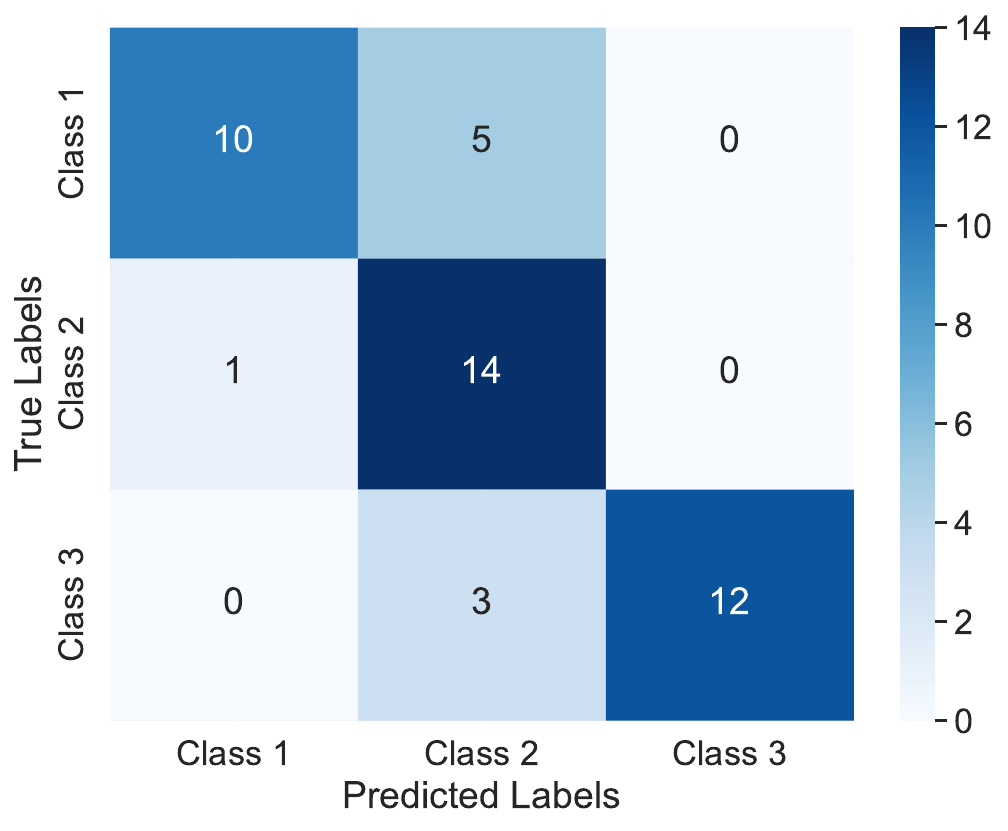}
  \caption{}
\end{subfigure}

\caption{Confusion matrix for the different classification models for the three class spiral dataset. Subfigure (a) represents the confusion matrix for the Q-RBF with 50 centres; subfigure (b) shows the C-RBF with 50 centres; subfigure (c) shows the MLP confusion matrix; and subfigure (d) shows the Gaussian SVM. It is clear that the RBF approaches perform better at classifying this data compared to the MLP and Gaussian SVM. This is highlighted in Table \ref{accuracy_table}. Note that the classes have different values due to the shuffling of the data, before the train:test split.}
\label{conf_mat}
\end{figure*}

\begin{figure*}[ht]

\begin{subfigure}{.5\linewidth}
  \includegraphics[width=\linewidth]{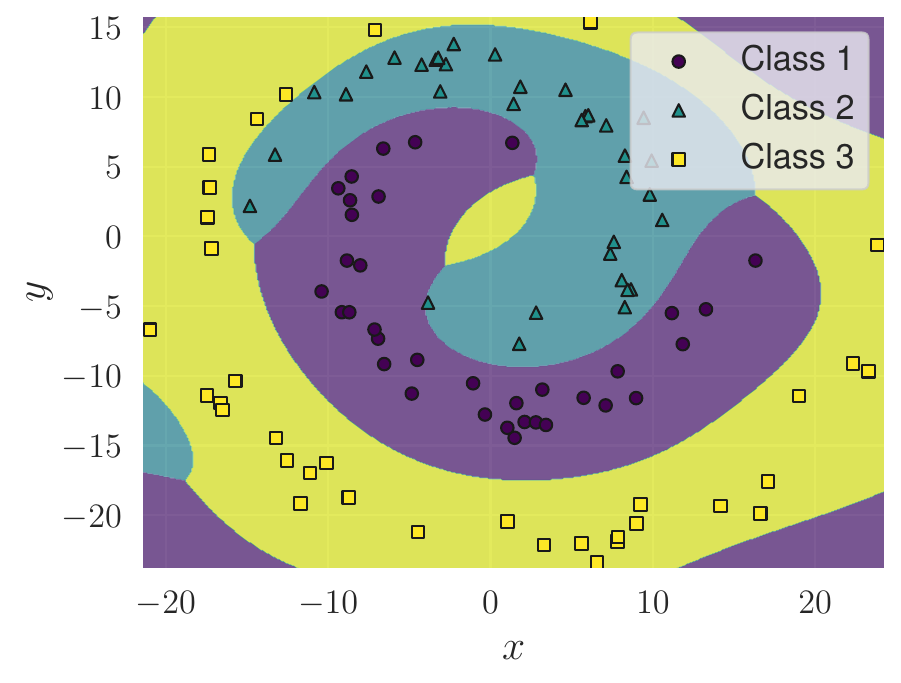}
  \caption{}
\end{subfigure}\hfill
\begin{subfigure}{.5\linewidth}
  \includegraphics[width=\linewidth]{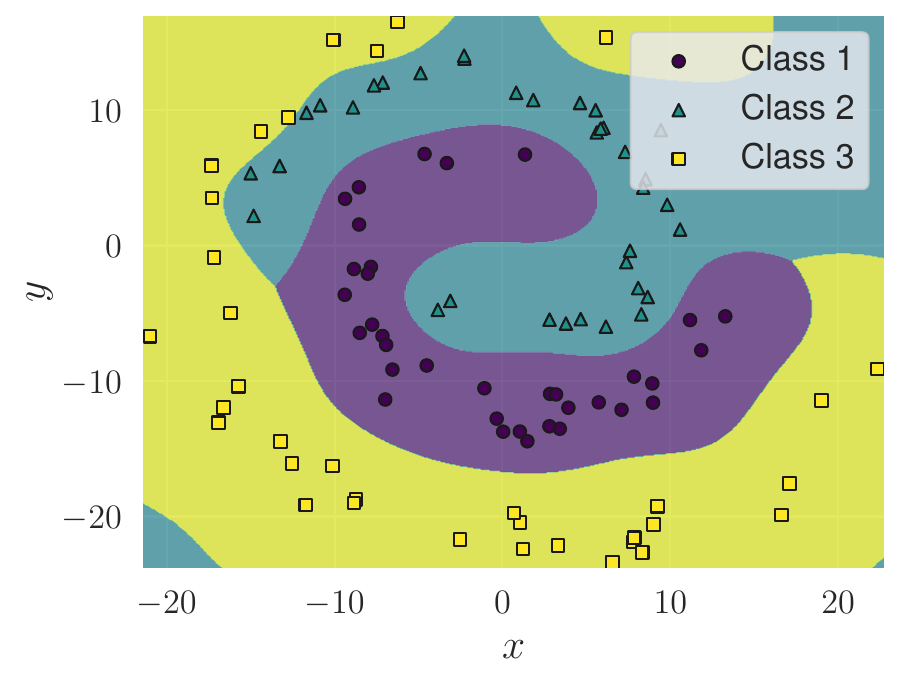}
  \caption{}
\end{subfigure}

\medskip 
\begin{subfigure}{.5\linewidth}
  \includegraphics[width=\linewidth]{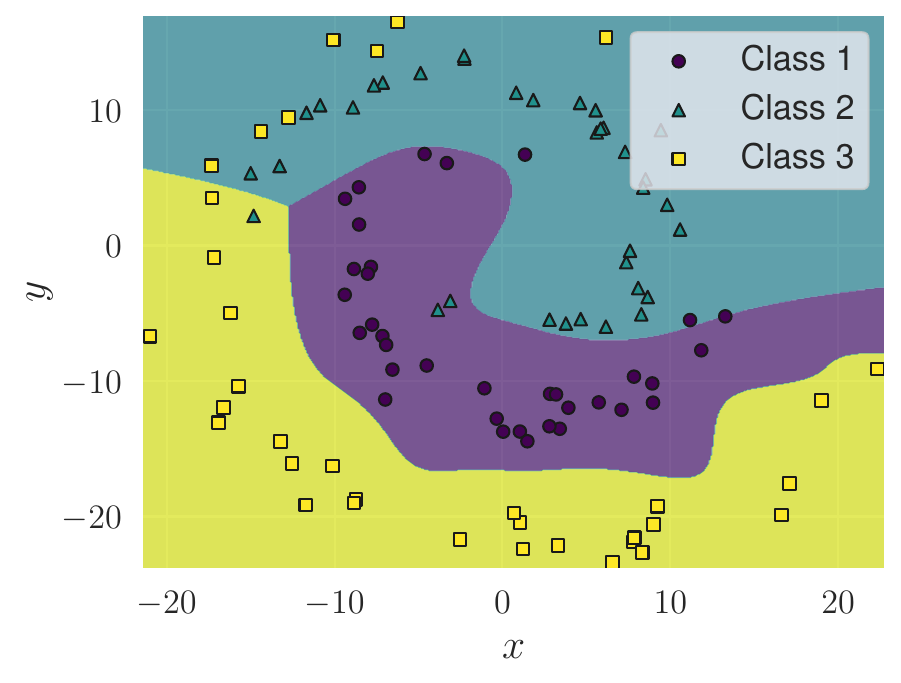}
  \caption{}
\end{subfigure}\hfill 
\begin{subfigure}{.5\linewidth}
  \includegraphics[width=\linewidth]{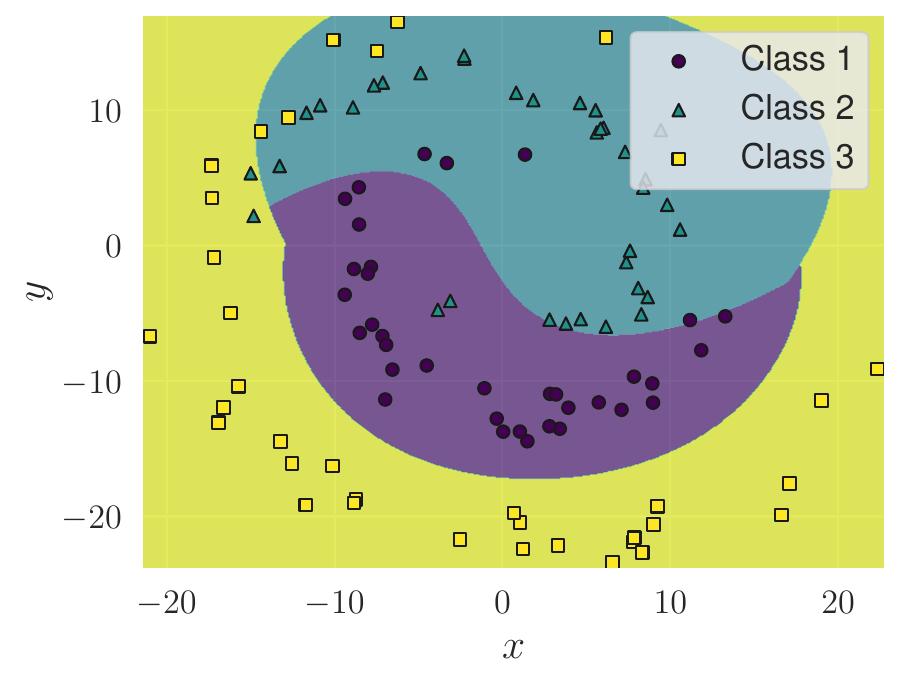}
  \caption{}
\end{subfigure}

\caption{Decision boundaries for the different classification models, where the subfigures are labelled as before: (a) Q-RBF; (b) C-RBF; (c) MLP; and (d) Gaussian SVM. This provides a pictorial representation highlighting which classes have been correctly identified by the algorithm by generating a grid of points over the feature space and using the trained models to predict over the grid.}
\label{class_fig}
\end{figure*}

Table 1 shows the mean squared error for the various models considered when interpolating. It is clear that when a sufficient number of centres are chosen, the quantum RBF performs comparatively well, against the classical RBF, SVM, and MLP. Given the consistency with which the quantum RBF interpolates out-of-sample data to a high accuracy, this suggests that this approach is a potentially feasible hybrid quantum-classical algorithm.

\section{Classification} \label{classification}
Given the success in interpolation, we now consider the quantum RBF's capability at performing classification tasks. Given interpolation and classification are inherently linked, this provides the quantum RBF with the opportunity to be used for more practical applications. Furthermore, as highlighted previously, the RBF network offers multiclass classification capability natively. This is distinctly different to the SVM which typically requires sequential operations in order to perform multiclass classification. 
\subsection{Spiral Dataset}

Spiral datasets are a popular benchmarking problem to evaluate the performance of machine learning models on non-linearly separable data. The standard dataset is a two-class problem, however the example considered here has an extra class adding further complexity. Mathematically, the spiral dataset is generated by the general equation for a point, $i$, on a spiral $(x_i,y_i)$
\begin{equation}
\label{spiral_eq}
    \begin{pmatrix}
        x^{(k)}_i \\ y^{(k)}_i
    \end{pmatrix}
    = r^{(k)}(\theta_i) \begin{pmatrix}
        \cos\theta_i \\ \sin\theta_i
    \end{pmatrix} + \begin{pmatrix}
        X_i \\ Y_i
    \end{pmatrix},
\end{equation}
where $r^{(k)}(\theta)$ is the radius of the spiral, $\theta_i$ is the angle between the points on the spiral and the x-axis, given by $\theta = 2\pi\sqrt{i}$, $X_i, Y_i \sim U(0,1)$, where $U(0,1)$ is a uniform distribution between 0 and 1, and $k$ represents the spiral class. The spiral dataset generated for this work has 3 classes, $k \in \{1,2,3\}$, each with 50 instances, as seen in Figure \ref{class_fig}. For each spiral $k \in \{1,2,3\}$ the radius function was defined as
\begin{equation}
    r^{(k)}(\theta) = 2\theta + a_k,
\end{equation}
where $a_k$ is the offset of each spiral. In this work $a_k$ were chosen as $\pi$, $-\pi$ and $4\pi$ for each class respectively. The generated data can be seen in Figure \ref{spiral dataset}. Once the data was generated, each class was shuffled to randomise the order of the classes. \newline 
The Q-RBF, C-RBF, SVM and MLP models were assessed for the generated 3-class spiral dataset, using a 70:30 train:test split. For the quantum and classical RBF networks, one-hot encoding was used to transform the spiral class targets into a binary format. This creates a column for each class where 1 indicates the class is present and 0 indicates it is not. Therefore for this spiral example, $m=105$ (for the training data), $q=3$ (due to the one-hot encoding), and $P=2$ (due to the $x,y$ dimensions). The number of centres chosen was 50, and these were randomly sampled for the Q-RBF network. For the C-RBF network, the centres were sampled using K-Means clustering, in the data range of $x$ and $y$.

\begin{table}
    \centering \caption{Accuracy of test data classifications for the spiral dataset} \label{spiral}
\begin{tabular}{ c | c }
\hline
Model & Accuracy \\
\hline \hline
 Q-RBF $j=50$ & 0.956  \\
 \hline
 Classical RBF $j=50$ & 0.956 \\
 \hline 
 Linear SVM & 0.622 \\
 \hline
 Gaussian SVM & 0.800 \\
 \hline
 MLP & 0.822
\end{tabular}
\label{accuracy_table}
\end{table}

\begin{figure}
    \centering
    \includegraphics[width=\linewidth]{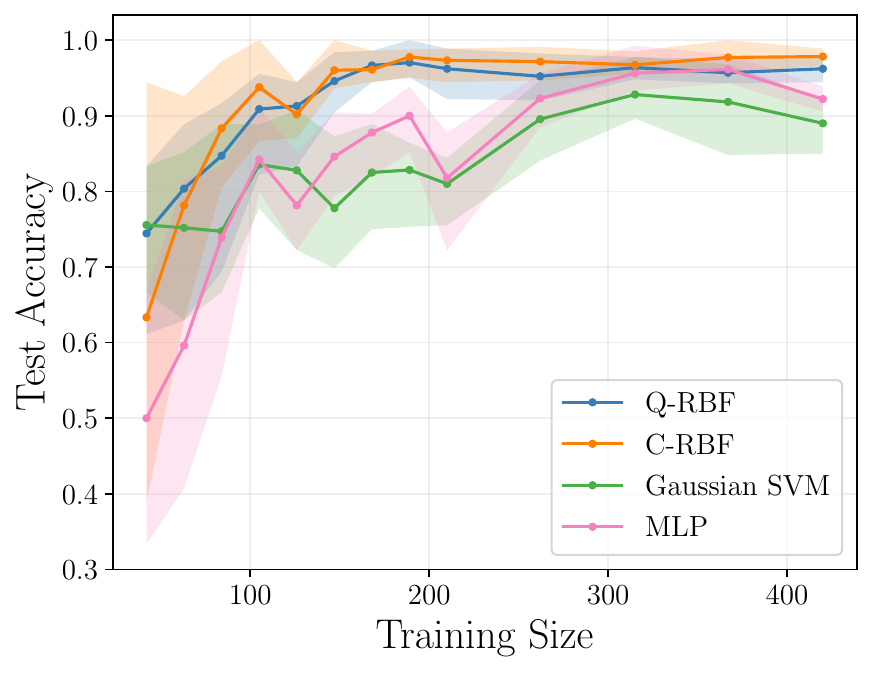}
    \caption{This figure demonstrates the increasing accuracy as a function of training data size for the spiral dataset, with a 70:30 train:test split. There are error bars in the RBF's due to the centres being sampled from a random distribution over the interval. Similar errors are introduced for the MLP and SVM.}
    \label{accuracy_training}
\end{figure}

To determine the accuracy for each of the different models, it is useful to understand how the accuracy changes as a function of the (relative) training data size. It is expected that as the training data is increased this would lead to the accuracy increasing for the given dataset. This is what is observed in Fig. \ref{accuracy_training}. It should be mentioned that whilst increasing the number of training datapoints may increase the accuracy on the training data (for a fixed number of overall datapoints), it would likely lead to worse accuracy on the test data as the fitted model would not be generalisable to other relevant datasets. However, increasing the number of overall datapoints, in addition to the number of training datapoints would likely lead to improved accuracy in the test data as the found model is not overfitting the data based on the training data.

\subsection{Iris Dataset}
The Iris dataset is a well-known classification dataset for machine learning algorithms, consisting of 3 classes of 50 instances each \citep{iris_dataset}. Each class corresponds to a type of Iris plant, setosa, virginica or versicolor, and there are 4 features for each sample, the length and width of the flower sepal and petals. The same models as those discussed in Section \ref{classification} are used for the classification on the Iris dataset. To ensure consistency, a 70:30 train:test split was also used for this dataset. For this Iris example, $m=105$ (for the training data), $q=3$ (due to the one-hot encoding and 3 classes of flower), and $P=4$ (due to number of features in the dataset). The number of centres chosen was 50 and found as before for the spiral dataset for the Q-RBF network and C-RBF network.

Table \ref{iris} shows the results of the Iris dataset classification. The results of the Q-RBF are clearly comparable with those of the MLP and the C-RBF models, achieving perfect accuracy.

\begin{table}
    \centering \caption{Accuracy of test data classifications for the Iris dataset} \label{iris}
\begin{tabular}{ c | c }
\hline
Model & Accuracy \\
\hline \hline
 Q-RBF $j=50$ & 1.0  \\
 \hline
 Classical RBF $j=50$ & 1.0 \\
 \hline 
 Linear SVM & 0.978 \\
 \hline
 Gaussian SVM & 0.978 \\
 \hline
 MLP & 1.0
\end{tabular}
\end{table}

\section{Discussion} \label{discussion}
Whilst this paper demonstrates the potential of a radial basis function where the data encoding has been performed using a quantum kernel approach, there are several key questions that remain unanswered. One significant issue is the need to understand how the algorithm performs in both noisy simulation, and on real quantum hardware \citep{QSVM_ion}. For simple classification problems, it should be feasible to achieve this on state-of-the-art NISQ devices, as long as the number of features is low. A key requirement will be the need to entangle multiple qubits in order to ensure the data encoding is faithfully represented on the quantum computer. Understanding the technical feasibility of this will be necessary to understand the near-term applications of a quantum-RBF approach. Given the importance of mapping each individual datapoint (and its respective features) to a point on the Bloch sphere, it will be necessary to have high fidelity quantum hardware in order to faithfully represent the data accurately. This will be crucial for determining the potential benefit of this approach.

Another technical issue that should be addressed is understanding any potential theoretical improvement from this approach. Given the study undertaken here is primarily empirical, it should be mathematically understood whether quantum-kernel based approaches offer any potential benefit over standard classical machine learning algorithms \citep{exp_conc, xphb-x2g4,data_qml}. As with every quantum-kernel based approach, there are questions related to how to optimally encode the data, along with how the encoding affects the overall result \citep{Q_data_encoding,PhysRevA.107.012422,PhysRevA.102.032420}. Understanding whether there are certain types of classification problem which require different encodings will be necessary for understanding any potential benefit which may arise when using quantum-based approaches for classification \citep{PRXQuantum.2.040321}.

There should also be rigorous benchmarking conducted testing this approach against other quantum kernel-based methods, along with state-of-the-art classical techniques \citep{10015720}. Optimising for hyperparameters has not been conducted in this paper (e.g. understanding the optimal choice for $\alpha$ in the Q-RBF network for a given problem), however this would be necessary to understand the model's full capacity. It should be also be fully understood how these quantum kernel-based models can be implemented in real-world scenarios \citep{Shadow_QML,QSVM_CFD,PRXQuantum.3.030101}.

\section{Conclusion}
This paper introduces a new implementation of the radial basis function network using quantum kernels, inspired by the standard approach of using quantum kernels in support vector machines. Several tests have been conducted highlighting the capability of this approach for interpolation and classification, along with comparisons against similar classical machine learning techniques. The methodology undertaken here demonstrates the proof of concept of the Q-RBF network algorithm, highlighting the need for further investigation into this approach. The next steps are to develop and implement this algorithm on real quantum hardware to further demonstrate its feasibility for near-term quantum computing applications.

\section{Acknowledgements}
The authors acknowledge stimulating discussions with Leo Bungay around the encoding of the data, David Lowe for providing background on the original development of the classical radial basis function network, and Gillian Marshall for reviewing the manuscript.

\bibliographystyle{ieeetr}
\bibliography{reference_file} 

@article{PhysRevA.107.032428,
  title = {Quantum kernel methods for solving regression problems and differential equations},
  author = {Paine, Annie E. and Elfving, Vincent E. and Kyriienko, Oleksandr},
  journal = {Phys. Rev. A},
  volume = {107},
  issue = {3},
  pages = {032428},
  numpages = {13},
  year = {2023},
  publisher = {American Physical Society},
  doi = {10.1103/PhysRevA.107.032428},
  url = {https://link.aps.org/doi/10.1103/PhysRevA.107.032428}
}

@article{Broomhead1988MultivariableFI,
  title={Multivariable Functional Interpolation and Adaptive Networks},
  author={David S. Broomhead and David Lowe},
  journal={Complex Syst.},
  year={1988},
  volume={2},
  url={https://api.semanticscholar.org/CorpusID:3686496}
}

@article{svm_ref,
  title={Support-vector networks},
  author={Corinna Cortes and Vladimir Vapnik },
  journal={Mach Learn},
  year={1995},
  volume={20},
  pages = {273},
  url={https://doi.org/10.1007/BF00994018}
}

@ARTICLE{991427,
  author={Chih-Wei Hsu and Chih-Jen Lin},
  journal={IEEE Transactions on Neural Networks}, 
  title={A comparison of methods for multiclass support vector machines}, 
  year={2002},
  volume={13},
  number={2},
  pages={415},
  doi={10.1109/72.991427}}

@ARTICLE{svm_ref1,
  author={Parkavi Sridhar and Parthiban Angamuthu},
  journal={Sci Rep}, 
  title={Enhancing image based classification for crop disease detection using a multiclass SVM approach with kernel comparison}, 
  year={2025},
  volume={15},
  pages={40055},
  doi={https://doi.org/10.1038/s41598-025-23568-w}}

@article{PhysRevLett.113.130503,
  title = {Quantum Support Vector Machine for Big Data Classification},
  author = {Rebentrost, Patrick and Mohseni, Masoud and Lloyd, Seth},
  journal = {Phys. Rev. Lett.},
  volume = {113},
  issue = {13},
  pages = {130503},
  numpages = {5},
  year = {2014},
  publisher = {American Physical Society},
  doi = {10.1103/PhysRevLett.113.130503},
  url = {https://link.aps.org/doi/10.1103/PhysRevLett.113.130503}
}

@article{natQSVM,
  title = {Supervised learning with quantum-enhanced feature spaces},
  author = {Vojtěch Havlíček and Antonio D. Córcoles and Kristan Temme and Aram W. Harrow and Abhinav Kandala and Jerry M. Chow and Jay M. Gambetta},
  journal = {Nature},
  volume = {567},
  pages = {209},
  year = {2019},
  doi = {10.1038/s41586-019-0980-2},
  url = {https://doi.org/10.1038/s41586-019-0980-2}
}

@article{log_map,
  title = {Simple mathematical models with very complicated dynamics},
  author = {Robert May},
  journal = {Nature},
  volume = {261},
  pages = {459},
  year = {1976},
  doi = {https://doi.org/10.1038/261459a0},
  url = {https://doi.org/10.1038/s41586-019-0980-2}
}

@article{kernel_class,
  title = {Unsupervised quantum machine learning for fraud detection},
  author = {Kyriienko, Oleksandr and Magnusson, Einar B.},
  journal = {arXiv:2208.01203},
  year = {2022},
  doi = {10.48550/arXiv.2208.01203},
  url = {https://doi.org/10.48550/arXiv.2208.01203}
}

@article{PhysRevLett.114.140504,
  title = {Experimental Realization of a Quantum Support Vector Machine},
  author = {Li, Zhaokai and Liu, Xiaomei and Xu, Nanyang and Du, Jiangfeng},
  journal = {Phys. Rev. Lett.},
  volume = {114},
  issue = {14},
  pages = {140504},
  numpages = {5},
  year = {2015},
  publisher = {American Physical Society},
  doi = {10.1103/PhysRevLett.114.140504},
  url = {https://link.aps.org/doi/10.1103/PhysRevLett.114.140504}
}

@article{qsvm_class_trapped_ion,
  title = {Quantum support vector machines for classification and regression on a trapped-ion quantum computer},
  author = {Teppei Suzuki and Takashi Hasebe and Tsubasa Miyazaki },
  journal = {Quantum Machine Intelligence},
  volume = {6},
  issue = {31},
  year = {2024},
  doi = {10.1007/s42484-024-00165-0},
  url = {https://doi.org/10.1007/s42484-024-00165-0}
}

@article{qsvm_class_photonic,
  title = {Experimental quantum-enhanced kernel-based machine learning on a photonic processor},
  author = {Zhenghao Yin and Iris Agresti and Giovanni de Felice and Douglas Brown and Alexis Toumi and Ciro Pentangelo and Simone Piacentini and Andrea Crespi and Francesco Ceccarelli and Roberto Osellame and Bob Coecke and Philip Walther},
  journal = {Nature Photonics},
  volume = {19},
  pages = {1020},
  year = {2025},
  doi = {10.1038/s41566-025-01682-5},
  url = {https://doi.org/10.1038/s41566-025-01682-5}
}

@article{qsvm_complexity,
  title = {The complexity of quantum support vector machines},
  author = {Gian Gentinetta1 and Arne Thomsen and David Sutter and Stefan Woerner},
  journal = {Quantum},
  volume = {8},
  pages = {1225},
  year = {2024},
  doi = {10.22331/q-2024-01-11-1225},
  url = {https://doi.org/10.22331/q-2024-01-11-1225}
}

@INPROCEEDINGS{374353,
  author={Lowe, D.},
  booktitle={Proceedings of 1994 IEEE International Conference on Neural Networks (ICNN'94)}, 
  title={Non-local radial basis functions for forecasting and density estimation}, 
  year={1994},
  volume={2},
  number={},
  pages={1197-1198},
  doi={10.1109/ICNN.1994.374353}}

@article{shakespeare,
  title = {Shakespeare vs. fletcher: A stylometric analysis by radial basis functions},
  author = {David Lowe and Robert Matthews},
  journal = {Comput Hum},
  volume = {29},
  pages = {449},
  year = {1995},
  doi = {10.1007/BF01829876},
  url = {https://doi.org/10.1007/BF01829876}
}

@ARTICLE{6795628,
  author={Bishop, Chris},
  journal={Neural Computation}, 
  title={Improving the Generalization Properties of Radial Basis Function Neural Networks}, 
  year={1991},
  volume={3},
  pages={579},
  doi={10.1162/neco.1991.3.4.579}}

@article{PhysRevD.104.076011,
  title = {Application of radial basis functions neural networks in spectral functions},
  author = {Zhou, Meng and Gao, Fei and Chao, Jingyi and Liu, Yu-Xin and Song, Huichao},
  journal = {Phys. Rev. D},
  volume = {104},
  pages = {076011},
  year = {2021},
  doi = {10.1103/PhysRevD.104.076011},
  url = {https://link.aps.org/doi/10.1103/PhysRevD.104.076011}
}

@article{TIPPING1998211,
author = {Michael E. Tipping and David Lowe},
title = {Shadow targets: A novel algorithm for topographic projections by radial basis functions},
journal = {Neurocomputing},
volume = {19},
pages = {211},
year = {1998},
doi = {10.1016/S0925-2312(97)00066-0},
url = {https://www.sciencedirect.com/science/article/pii/S0925231297000660}
}

@article{
doi:10.1126/science.abn7293,
author = {Hsin-Yuan Huang  and Michael Broughton  and Jordan Cotler  and Sitan Chen  and Jerry Li  and Masoud Mohseni  and Hartmut Neven  and Ryan Babbush  and Richard Kueng  and John Preskill  and Jarrod R. McClean },
title = {Quantum advantage in learning from experiments},
journal = {Science},
volume = {376},
pages = {1182},
year = {2022},
doi = {10.1126/science.abn7293},
URL = {https://www.science.org/doi/abs/10.1126/science.abn7293},
}

@article{
google2,
author = {Google},
title = {Observation of constructive interference at the edge of quantum ergodicity},
journal = {Nature},
volume = {646},
pages = {825},
year = {2025},
doi = {10.1038/s41586-025-09526-6},
URL = {https://doi.org/10.1038/s41586-025-09526-6},
}

@article{
google1,
author = {Arute, F. and Arya, K. and Babbush, R. and others},
title = {Quantum supremacy using a programmable superconducting processor.},
journal = {Nature},
volume = {574},
pages = {505},
year = {2019},
doi = {10.1038/s41586-019-1666-5},
URL = {https://doi.org/10.1038/s41586-019-1666-5},
}

@article{
QML_rev,
author = {Jacob Biamonte and Peter Wittek and Nicola Pancotti and Patrick Rebentrost and Nathan Wiebe and Seth Lloyd },
title = {Quantum machine learning},
journal = {Nature},
volume = {549},
pages = {195},
year = {2017},
doi = {10.1038/nature23474},
URL = {https://doi.org/10.1038/nature23474},
}

@article{iris_dataset,
author = {Fisher, R. A.},
title = {THE USE OF MULTIPLE MEASUREMENTS IN TAXONOMIC PROBLEMS},
journal = {Annals of Eugenics},
volume = {7},
pages = {179},
doi = {https://doi.org/10.1111/j.1469-1809.1936.tb02137.x},
year = {1936},
url = {https://onlinelibrary.wiley.com/doi/abs/10.1111/j.1469-1809.1936.tb02137.x},
eprint = {https://onlinelibrary.wiley.com/doi/pdf/10.1111/j.1469-1809.1936.tb02137.x}
}

@article{PhysRevA.102.042418,
  title = {Data classification by quantum radial-basis-function networks},
  author = {Shao, Changpeng},
  journal = {Phys. Rev. A},
  volume = {102},
  issue = {4},
  pages = {042418},
  numpages = {9},
  year = {2020},
  publisher = {American Physical Society},
  doi = {10.1103/PhysRevA.102.042418},
  url = {https://link.aps.org/doi/10.1103/PhysRevA.102.042418}
}

@article{ZHOU2025129254,
title = {Enhancement of radial basis function model via quantum kernel estimation},
journal = {Journal of Mathematical Analysis and Applications},
volume = {547},
pages = {129254},
year = {2025},
issn = {0022-247X},
doi = {https://doi.org/10.1016/j.jmaa.2025.129254},
url = {https://www.sciencedirect.com/science/article/pii/S0022247X25000356},
author = {Xiaojian Zhou and Meng Zhang and Qi Cui and Ting Jiang}
}

@article{Huggins_2019,
doi = {10.1088/2058-9565/aaea94},
url = {https://doi.org/10.1088/2058-9565/aaea94},
year = {2019},
publisher = {IOP Publishing},
volume = {4},
number = {2},
pages = {024001},
author = {Huggins, William and Patil, Piyush and Mitchell, Bradley and Whaley, K Birgitta and Stoudenmire, E Miles},
title = {Towards quantum machine learning with tensor networks},
journal = {Quantum Science and Technology}
}

@article{PARK2020126422,
title = {The theory of the quantum kernel-based binary classifier},
journal = {Physics Letters A},
volume = {384},
pages = {126422},
year = {2020},
issn = {0375-9601},
doi = {https://doi.org/10.1016/j.physleta.2020.126422},
url = {https://www.sciencedirect.com/science/article/pii/S0375960120302541},
author = {Daniel K. Park and Carsten Blank and Francesco Petruccione}
}

@article{Shadow_QML,
title = {Shadows of quantum machine learning},
journal = {Nature Communications },
volume = {15},
pages = {5676},
year = {2024},
doi = {https://doi.org/10.1038/s41467-024-49877-8},
url = {https://doi.org/10.1038/s41467-024-49877-8},
author = {Sofiene Jerbi and Casper Gyurik and Simon C. Marshall and Riccardo Molteni and Vedran Dunjko},
}

@article{QSVM_ion,
title = {Quantum support vector machines for classification and regression on a trapped-ion quantum computer},
journal = {Quantum Mach. Intell.},
volume = {6},
pages = {31},
year = {2024},
doi = {https://doi.org/10.1007/s42484-024-00165-0},
url = {https://doi.org/10.1007/s42484-024-00165-0},
author = {Teppei Suzuki and Takashi Hasebe and Tsubasa Miyazaki },
}

@article{
QSVM_CFD,
author = {Xi-Jun Yuan  and Zi-Qiao Chen  and Yu-Dan Liu  and Zhe Xie  and Ying-Zheng Liu  and Xian-Min Jin  and Xin Wen  and Hao Tang },
title = {Quantum Support Vector Machines for Aerodynamic Classification},
journal = {Intelligent Computing},
volume = {2},
number = {},
pages = {0057},
year = {2023},
doi = {10.34133/icomputing.0057},
URL = {https://spj.science.org/doi/abs/10.34133/icomputing.0057}
}

@article{
exp_conc,
author = {Supanut Thanasilp and Samson Wang and M. Cerezo and Zoë Holmes },
title = {Exponential concentration in quantum kernel methods},
journal = {Nature Communications},
volume = {15},
pages = {5200},
year = {2024},
doi = {https://doi.org/10.1038/s41467-024-49287-w},
URL = {https://doi.org/10.1038/s41467-024-49287-w}
}

@article{xphb-x2g4,
  title = {Neural quantum kernels: Training quantum kernels with quantum neural networks},
  author = {Rodriguez-Grasa, Pablo and Ban, Yue and Sanz, Mikel},
  journal = {Phys. Rev. Res.},
  volume = {7},
  issue = {2},
  pages = {023269},
  numpages = {16},
  year = {2025},
  publisher = {American Physical Society},
  doi = {10.1103/xphb-x2g4},
  url = {https://link.aps.org/doi/10.1103/xphb-x2g4}
}

@article{Q_PCA,
title = {Quantum principal component analysis},
journal = {Nature Phys},
volume = {10},
pages = {631},
year = {2014},
issn = {0375-9601},
doi = {https://doi.org/10.1038/nphys3029},
url = {https://doi.org/10.1038/nphys3029},
author = {Lloyd, S. and Mohseni, M. and Rebentrost, P}
}

@article{Q_data_encoding,
title = {Quantum data encoding: a comparative analysis of classical-to-quantum mapping techniques and their impact on machine learning accuracy},
journal = {EPJ Quantum Technol.},
volume = {11},
pages = {72},
year = {2024},
doi = {https://doi.org/10.1140/epjqt/s40507-024-00285-3},
url = {https://doi.org/10.1140/epjqt/s40507-024-00285-3},
author = {Minati Rath and Hema Date }
}

@article{data_qml,
title = {Power of data in quantum machine learning},
journal = {Nature Communications },
volume = {12},
pages = {2631},
year = {2021},
doi = {https://doi.org/10.1038/s41467-021-22539-9},
url = {https://doi.org/10.1038/s41467-021-22539-9},
author = {Hsin-Yuan Huang and Michael Broughton and Masoud Mohseni and Ryan Babbush and Sergio Boixo and Hartmut Neven and Jarrod R. McClean }
}

@article{PhysRevA.102.032420,
  title = {Robust data encodings for quantum classifiers},
  author = {LaRose, Ryan and Coyle, Brian},
  journal = {Phys. Rev. A},
  volume = {102},
  issue = {3},
  pages = {032420},
  numpages = {24},
  year = {2020},
  publisher = {American Physical Society},
  doi = {10.1103/PhysRevA.102.032420},
  url = {https://link.aps.org/doi/10.1103/PhysRevA.102.032420}
}

@article{PhysRevA.107.012422,
  title = {Exponential data encoding for quantum supervised learning},
  author = {Shin, S. and Teo, Y. S. and Jeong, H.},
  journal = {Phys. Rev. A},
  volume = {107},
  issue = {1},
  pages = {012422},
  numpages = {20},
  year = {2023},
  publisher = {American Physical Society},
  doi = {10.1103/PhysRevA.107.012422},
  url = {https://link.aps.org/doi/10.1103/PhysRevA.107.012422}
}

@article{Q_Reinf_learn,
title = {Experimental quantum speed-up in reinforcement learning agents},
journal = {Nature},
volume = {591},
pages = {229},
year = {2021},
doi = {https://doi.org/10.1038/s41586-021-03242-7},
url = {https://doi.org/10.1038/s41586-021-03242-7},
author = {V. Saggio and B. E. Asenbeck and A. Hamann and T. Strömberg and P. Schiansky and V. Dunjko and N. Friis and N. C. Harris and M. Hochberg and D. Englund and S. Wölk and H. J. Briegel and P. Walther}
}

@article{Q_CNN,
title = {Quantum convolutional neural networks},
journal = {Nature Phys},
volume = {15},
pages = {1273},
year = {2019},
doi = {https://doi.org/10.1038/s41567-019-0648-8},
url = {https://doi.org/10.1038/s41567-019-0648-8},
author = {Cong, I. and Choi, S. and Lukin, M.D.}
}

@ARTICLE{10015720,
  author={Simões, Ricardo Daniel Monteiro and Huber, Patrick and Meier, Nicola and Smailov, Nikita and Füchslin, Rudolf M. and Stockinger, Kurt},
  journal={IEEE Access}, 
  title={Experimental Evaluation of Quantum Machine Learning Algorithms}, 
  year={2023},
  volume={11},
  pages={6197},
  doi={10.1109/ACCESS.2023.3236409}}

@article{PRXQuantum.3.030101,
  title = {Is Quantum Advantage the Right Goal for Quantum Machine Learning?},
  author = {Schuld, Maria and Killoran, Nathan},
  journal = {PRX Quantum},
  volume = {3},
  issue = {3},
  pages = {030101},
  numpages = {13},
  year = {2022},
  publisher = {American Physical Society},
  doi = {10.1103/PRXQuantum.3.030101},
  url = {https://link.aps.org/doi/10.1103/PRXQuantum.3.030101}
}

@article{PRXQuantum.2.040321,
  title = {Generalization in Quantum Machine Learning: A Quantum Information Standpoint},
  author = {Banchi, Leonardo and Pereira, Jason and Pirandola, Stefano},
  journal = {PRX Quantum},
  volume = {2},
  issue = {4},
  pages = {040321},
  numpages = {21},
  year = {2021},
  publisher = {American Physical Society},
  doi = {10.1103/PRXQuantum.2.040321},
  url = {https://link.aps.org/doi/10.1103/PRXQuantum.2.040321}
}

@article{VQA,
title = {Variational quantum algorithms},
journal = {Nature Rev. Phys},
volume = {3},
pages = {625},
year = {2021},
doi = {https://doi.org/10.1038/s42254-021-00348-9},
url = {https://doi.org/10.1038/s42254-021-00348-9},
author = {M. Cerezo and A. Arrasmith and R. Babbush and S. Benjamin and Suguru Endo and Keisuke Fujii and Jarrod R. McClean and Kosuke Mitarai and Xiao Yuan and Lukasz Cincio and Patrick J. Coles}
}

@article{Mele2024introductiontohaar,
  doi = {10.22331/q-2024-05-08-1340},
  url = {https://doi.org/10.22331/q-2024-05-08-1340},
  title = {Introduction to {H}aar {M}easure {T}ools in {Q}uantum {I}nformation: {A} {B}eginner's {T}utorial},
  author = {Mele, Antonio Anna},
  journal = {{Quantum}},
  issn = {2521-327X},
  publisher = {{Verein zur F{\"{o}}rderung des Open Access Publizierens in den Quantenwissenschaften}},
  volume = {8},
  pages = {1340},
  month = may,
  year = {2024}
}

@article{PRXQuantum.5.010309,
  title = {Entanglement Transitions in Unitary Circuit Games},
  author = {Morral-Yepes, Ra\'ul and Smith, Adam and Sondhi, S.L. and Pollmann, Frank},
  journal = {PRX Quantum},
  volume = {5},
  issue = {1},
  pages = {010309},
  numpages = {18},
  year = {2024},
  publisher = {American Physical Society},
  doi = {10.1103/PRXQuantum.5.010309},
  url = {https://link.aps.org/doi/10.1103/PRXQuantum.5.010309}
}

@misc{classicalRBF,
title = {{Building Your First Radial Basis Function Network in Python (or R)}},
year = {2025},
note = {Online; accessed 20 December 2025},
author = {Riswanto, U},
url = {https://ujangriswanto08.medium.com/building-your-first-radial-basis-function-network-in-python-or-r-3766c5da1ee7}
}

\end{document}